\documentclass{JAIS}

\journal{JAIS-ID}
\vol{2022}

\received{xx January 2018}
\published{xx March 2018}

\def\be{\begin{equation}}
\def\ee{\end{equation}}
\def\bea{\begin{eqnarray}}
\def\eea{\end{eqnarray}}

\usepackage{caption}
\usepackage{lineno}
\usepackage{amsmath}
\usepackage{mathtools}
\usepackage{xcolor}
\usepackage{soul}
\articletype{}

\begin{document}

\title{B2G4: A synthetic data pipeline for the integration of Blender models in Geant4 simulation toolkit}

\author{Angel Bueno,\auno{1} Felix Sattler,\auno{1} Maximilian Perez Prada,\auno{1}, Maurice Stephan\auno{1}, Sarah Barnes\auno{1}}
\address{$^1$German Aerospace Center (DLR),\\ Institute for the Protection of Maritime Infrastructures, Germany
\\
\vspace{0.2cm}
Corresponding author: Angel Bueno\\
Email address: angel.bueno@dlr.de
}

\begin{abstract}
The correctness and precision of particle physics simulation software, such as Geant4, is expected to yield results that closely align with real-world observations or well-established theoretical predictions. Notably, the accuracy of these simulated outcomes is contingent upon the software's capacity to encapsulate detailed attributes, including its prowess in generating or incorporating complex geometrical constructs. While the imperatives of precision and accuracy are essential in these simulations, the need to manually code highly detailed geometries emerges as a salient bottleneck in developing software-driven physics simulations. This research proposes Blender-to-Geant4 (B2G4), a modular data workflow that utilizes Blender to create 3D scenes, which can be exported as geometry input for Geant4. B2G4 offers a range of tools to streamline the creation of simulation scenes with multiple complex geometries and realistic material properties. Here, we demonstrate the use of B2G4 in a muon scattering tomography application to image the interior of a sealed steel structure. The modularity of B2G4 paves the way for the designed scenes and tools to be embedded not only in Geant4, but in other scientific applications or simulation software.
\end{abstract}

\maketitle


\section{Introduction}
\label{sec:intro}


\noindent Muon tomography, or MT, leverages muons as a non-invasive imaging technique to visualize macroscopic three-dimensional volumes. MT harnesses two main interaction patterns to investigate a volume of interest (VOI): absorption and scattering~\cite{bonechiyandrea}. The absorption pattern is centered around observing the attenuation of muons as they penetrate the VOI and comparing with the open-sky expectation. The scattering pattern, also known as muon scattering tomography (MST), focuses on quantifying the deflection angles of muons as they traverse the VOI. The deflections depend on the physical properties of the material traversed by the muons and thus serve as the bedrock for tomographic reconstruction. The canonical application of MST aims to quantify material densities in concealed or obscured setups where direct inspection would be costly or not feasible, such as border security applications~\cite{sarah2023}, infrastructure monitoring~\cite{thompson2020muon}, non-destructive industrial testing~\cite{arbol2019non} or nuclear fuel cask surveillance~\cite{poulson2019application}.\\

\noindent Current procedures for MT simulation are often implemented in Geant4~\cite{agostinelli2003geant4}, a powerful particle physics simulation software that accurately models particle propagation and their complex interactions with matter within a well-defined setting. Geant4 encapsulates the entire simulation in a single volume, namely the world volume, in which the physics processes, the geometrical constructs of objects (so called solids), detectors, and VOIs are specified. The geometrical definition of a VOI and its solids entails native function calls in C++ to the Geant4 application programming interface (API) in order to delineate the VOI and the solid properties. The advanced geometry descriptors within the Geant4 API offer a versatile implementation for constructing approximate and realistic 3D environments and, thus, permit the precise examination and analysis of muon interactions and their subsequent effects \cite{maxi}. However, creating highly detailed VOIs in Geant4 can be challenging and time-consuming, as it often requires manual coding and much expertise. Therefore, established coding approaches in Geant4 fall short when it comes to efficiently designing and importing complex scenes. This limitation can hinder researchers needing rapid and smooth scene creation and exportation for validating their simulation projects.\\

\noindent This research proposes Blender-to-Geant4 (B2G4), a novel framework that synergizes the power of Blender's 3D modeling tools \cite{blender3D} to enhance scene design and simulations in Geant4. Blender offers a versatile, user-friendly interface to generate intricate geometries, leveraging the power of drag-and-drop placement, solid modeling precision, shape variance randomization, and intuitive material assignments. The suite of functionalities defined in B2G4 transplants the designed scene in Blender into a readable format for Geant4, enabling the automated export and import of scenes with minimal manual effort. The B2G4 framework streamlines the development of simulation scenes in Geant4 for a diverse range of applications. This paper highlights one example workflow, including generating and analyzing detailed scenes for muon tomography.

\section{Related work}
\label{sec:related}

\begin{figure}
    \centering
    \includegraphics[width=\textwidth,height=6.33cm]{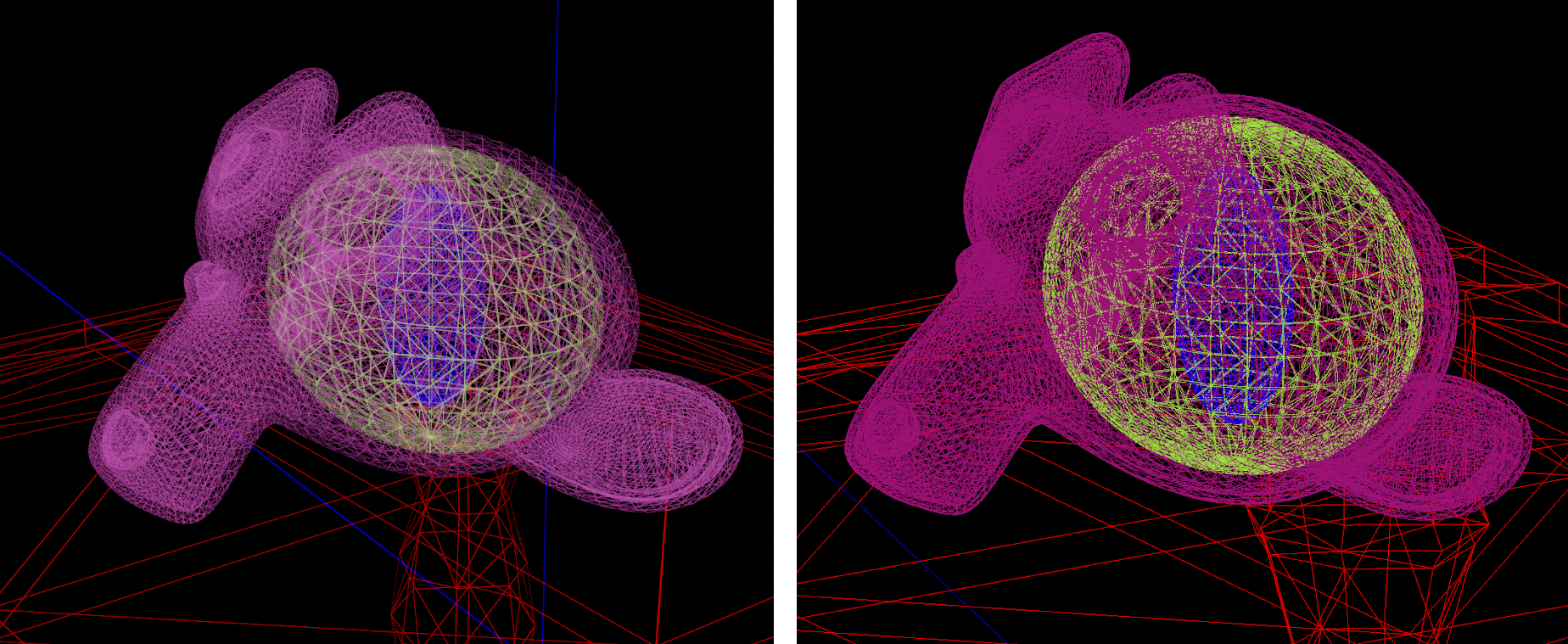}
    \caption{Scene setup with three solids in Blender (left) and Geant4 (right). The plastic monkey head (pink) with two ellipsoids (yellow, blue) of lead and uranium inside sits on top of a metallic surface (red) in a cuboid world (blue). The materials used are clearly distinguishable in both Blender and Geant4, demonstrating the finesse of B2G4 in solid and material definition.}
    \label{fig:figure1}
\end{figure}

Scene design in Geant4 with multiple solids is labor-intensive due to manual coding and iterative error rectifications. To speed up the scene design phase, several procedures have been suggested. The \textit{Geometry Description Markup Language} (GDML) was proposed by \cite{gdml2006}. The GDML language offers an XML-based schema designed to formulate geometry hierarchies in Geant4, emphasizing the reusability of geometric components to streamline the creation of solids in a simulation scene for Geant4. However, GDML entails manual manipulation of text files and sophisticated XML parsers for integration with Geant4, and notably, it lacks any form of visual feedback mechanism. An alternative approach presented by \cite{Constantin_2010} entails transferring 3D CAD data into GDML through proprietary 3D software like FastRad. This solution introduces the necessity of proficiency in commercial 3D CAD software.\\

\noindent Over the past years, the research community in Geant4 has shifted toward open software solutions. Research work by \cite{PINTO2019150} outlined an open-source workflow to convert STEP files to GDML, but the sophistication of the workflow requires manual coding and is prone to parsing errors across file formats. Proposed software by \cite{Poole2012} advocates integrating 3D models into Geant4 by converting standard 3D files to Geant4 code using the TesselatedFacets feature in Geant4. Though this software is a well-established method to export 3D models into Geant4, it does not extend this capability to entire 3D scenes; that is, the geometrical attributes of each solid in a scene, their positions, and logical hierarchies in Geant4 must be manually coded one by one. Further, when adding or modifying a 3D scene, these prior workflows require compilation in Geant4.\\

\noindent Our approach posits a newer direction in scene design for Geant4 using Blender, an open-source 3D modeling suite that is accessible, well-documented, and provides a Python interface to program external addons. Figure \ref{fig:figure1} depicts a scene design setup in Blender (left) and the exported scene in Geant4 (right) using our proposed B2G4 software. Notice that B2G4 permits the design of complete scenes with multiple solids, with precise geometry and material assignments. Further, B2G4 is open source (no proprietary tools are required) and provides enhanced visual feedback with easy scene editing. The dynamical scene exportation in B2G4 allows new scenes to be loaded during run time without the need for re-compilation of Geant4, a distinction crucial in scene design. In the following sections, we describe the working principles of B2G4 and explore its application in the context of muon scattering tomography.

\section{B2G4 description}
\label{sec:B2G4_description}

The standard B2G4 workflow consists of several stages: scene creation, parsing, saving, exporting into Geant4, and data interpretation. Figure 2 illustrates the complete workflow of B2G4, adapted for muon tomography data analysis. Note that these stages are grouped into three main modules: the \textit{B2G4 scene exporter}, the \textit{B2G4 data interpreter}, and the \textit{B2G4 muon tomography} module. After the data interpretation step by B2G4, the post-processed output is forwarded to specific use cases, here, muon tomography data analysis. The implementation of B2G4 into three independent modules enables easy scaling of the scene design process, from local prototyping in Blender to the simulation of randomized scenes at scale with loadable run-times in Geant4.\\

\begin{figure}
    \centering
    \includegraphics[width=\textwidth, height=4.3cm]{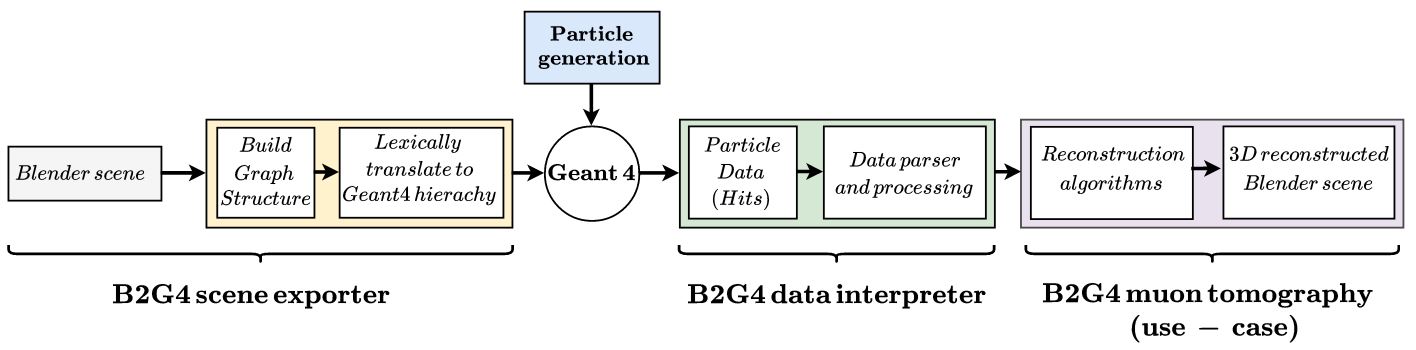}
    \caption{\textbf{Blender-to-Geant4 (B2G4) workflow:} B2G4 offers a range of tools to bridge Blender and Geant4, with Python scripts for Blender export and a C++ library for Geant4 imports. The \textit{B2G4 scene exporter} creates a graph structure that mimics the hierarchy of Geant4 scene structures, allowing the Blender scene to be translated into a format that Geant4 can read. The \textit{B2G4 data interpreter} processes particle data files produced by Geant4 simulation. The \textit{B2G4 muon tomography} framework then reconstructs the Blender scene using processed muon hits. The B2G4 modules are standalone, allowing users to employ this end-to-end workflow tailored to their specific simulation requirements in Geant4.}
    \label{fig:figure2}
\end{figure}

\noindent \textbf{B2G4 scene exporter:} The main module for scene creation and exportation is the \textit{B2G4 scene exporter}. This standalone module contains the internal scripts for the creation and saving of the scene in Blender, and the parser to Geant4: 

\begin{enumerate}
    \item \textbf{Scene creation:} B2G4 follows the code conventions of Geant4 to define the geometric attributes of the object being modeled, that is, each solid object in Geant4 is created by describing its attributes with reference to a logical volume, including shape and physical properties. Hence, the solid design in B2G4 maintains a one-to-one correspondence with the mother-daughter logical relationship in Geant4. This logical hierarchy is implemented using the \textit{Collection} module in Blender, a logical data container that permits the grouping of 3D solids for easier manipulation and organization of 3D scenes. The material assignment is done via the \textit{Material Properties} modifier in Blender. The available materials in B2G4 have been ported from the Geant4 material database and include pure elements, NIST compounds, biological, space, bio-chemical, high-energy physics, and nuclear materials. B2G4 adopts the ionization potential to aid in material differentiation and enhance visual clarity in analysis through contrasting colors. Every material category in B2G4 has a specific color map based on the normalized ionization potential of the materials specified in the Geant4 material database. We standardize the ionization potentials of materials in a specific category by comparing them to the minimum and maximum ionization potential. In cases where two materials have the same potential, we use their density to create a color offset for better visual separation. The normalized values are then converted to a color map. Figure~\ref{fig:figure1} depicts this color scheme for different materials for a monkey head lying on a metallic surface. The surface is $G4\_STAINLESS-STEEL$, whereas the monkey head is $G4\_Plastic$. The monkey brain is $G4\_Pb$, and the small cylindrical object inside is $G4\_U$.\\
    
    \item \textbf{Scene saving:} Once the scene has been designed, B2G4 allocates, for each \textit{Collection} in the scene, the mesh of the imported 3D solid, the position, and the assigned material. To achieve that, a dynamic graph is built by traversing the specified \textit{Collections} items. B2G4 defines a graph $G=(N, E)$, whose nodes $N$ contain a \textit{Collection} with the solid created and their attributes. The edges $E$ in this graph preserve the hierarchical logical relationship mother-daughter in Geant4, with the root node, the $world\,volume$. This graph $G$ serves as the basis for the exportation file required for Geant4. Next, B2G4 traverses and accesses the created graph hierarchy $G$ within the Blender environment. For every node (or \textit{Collection}) $N$ within the graph hierarchy $G$, relevant data is extracted to outline a logical volume, establish physical placement, and define a solid conforming to the native syntax of Geant4. After all the nodes are traversed, the resulting geometry tree encapsulates the data hierarchy of Geant4 and is stored as a JSON file. This file mirrors the Geant4 layout and incorporates pointers to individual solids. Each unique solid undergoes a singular export process to minimize redundancy, being saved as a Polygon File Format (PLY). B2G4 employs the PLY format for object exportation to establish a standardized 3D asset library, given its widespread recognition and adoption as an open-source file standard for sharing 3D triangular mesh data. This approach seeks to ensure compatibility across Geant4 applications.\\
    
    \item \textbf{Scene importing:} Here, we describe the process of importing the JSON and PLY files into Geant4. Given that the structure of the JSON file adheres to Geant4's scene hierarchy, parsing the JSON data to its corresponding Geant4 objects becomes rapid and efficient. To allow dynamic loading of a scene during run-time of Geant4, we resort to a single header file that loads and parses both, the JSON and the PLY files. This header file needs to be included only once in the standard Geant4 compilation process, and no external dependencies are required.\\
    
    For each JSON entry, the required data is systematically extracted. When a reference to a solid appears, the associated PLY files are loaded and constructed with \textit{G4TesselatedSolid}. It is important to note that every solid is created only \textit{once} to save computational overhead, even though it could be referenced several times. The logical volumes (\textit{G4LogicalVolume}) are created by parsing the required data from the JSON file and looking up the referenced \textit{G4Solid} from the already loaded \textit{G4TesselatedSolids}. The resulting \textit{G4LogicalVolume} is stored in an array. The creation of \textit{G4PhysicalVolume} objects resort to a similar association of the correct transformation data from the JSON file and logical volume, based on the name identifier. 
\end{enumerate}

\begin{figure}
    \centering
    \includegraphics[width=\textwidth, height=10.75cm]{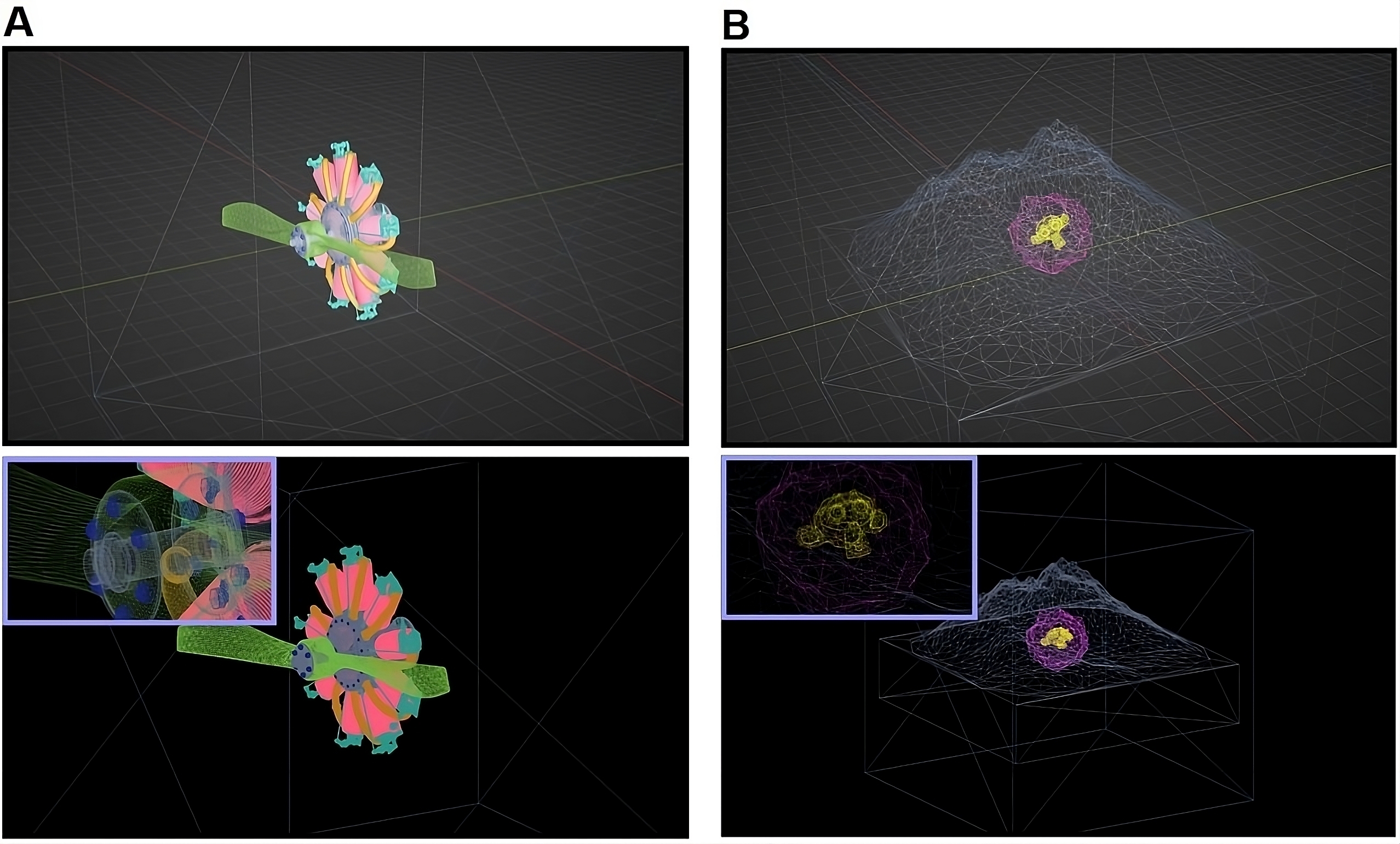}
    \caption{\textbf{A B2G4 example of scene design and parsing from Blender into Geant4:} the top portion showcases the solid designed with Blender, while the bottom portion displays its equivalent in Geant4. An inset between the two images highlights the extensive details within each solid, precisely transferred from Blender into Geant4. (A) depicts an aircraft engine with highly detailed bolts and rotatory blades (B) a mountain with a cave and a golden monkey head within. The ionization potential colormap in B2G4 eases the identification of each component in a scene. }
    \label{fig:figure3}
\end{figure}

\noindent B2G4 can import scenes of various scales with ease, from intricate designs to complex landscapes. Figure~\ref{fig:figure3} depicts two examples of scene design and exportation process using B2G4: a rotatory aircraft engine (Figure~\ref{fig:figure3}.a) and a mountain with a golden monkey head (Figure~\ref{fig:figure3}.b). Each solid in the scene is designed using the $Collection$ to maintain the one-to-one correspondence with the mother-daughter logical relationship in Geant4, including the $world\,volume$. The scene of the rotatory aircraft contains 1.5 million triangles, whereas the scene of the mountain scene contains 4487 triangles. Both are saved in 15.2 and 0.05 seconds, respectively. Once the scene have been exported to Geant4, users can implement their custom simulation setup.\\

\begin{figure*}[!ht]
    \centering
    \includegraphics[width=\textwidth, height=11.80cm]{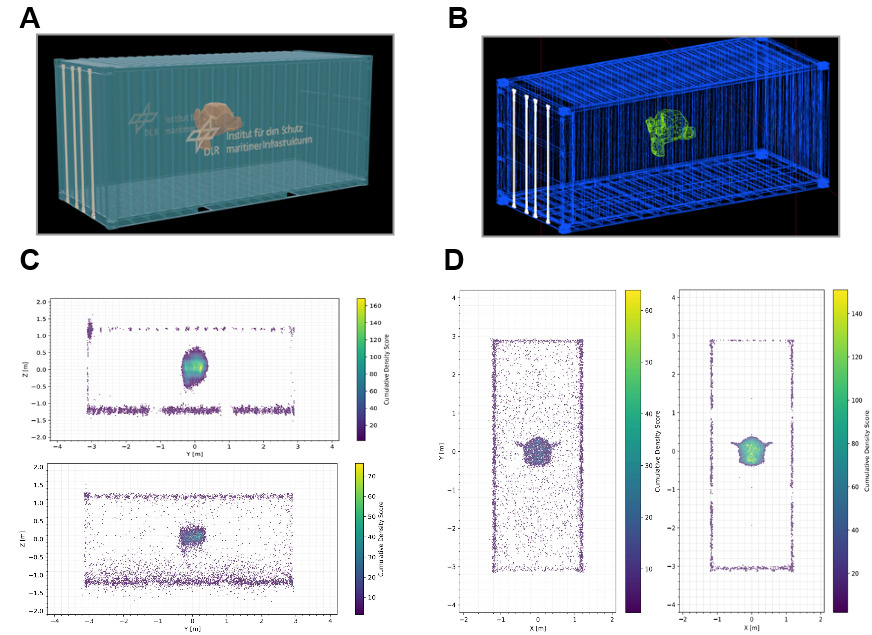}
    \caption{\textbf{Example of muon tomography:} A scene with a sealed steel container hiding a monkey head; (A) scene in Blender, with textures (B) exported scene in Geant4; (C) side view of the reconstruction in ASR (upper) and PoCA (lower); (D) top-view of the reconstruction with ASR (right) and PoCA (left).}
    \label{fig:figure4}
\end{figure*}

\noindent \textbf{Particle generation with Geant4:} The generation of particles is always customized to suit the Geant4 scientific application. This work uses \textit{Cosmic-ray Shower Library (CRY)}\cite{cry}, a well-known software, to generate cosmic ray air shower particles. The simulation data is created by recording details of the interactions of atmospheric muons with detector planes. The configuration of the detectors, including their thickness and materials, must be outlined. In this context, we operate under the assumption of ideal detectors, that is, no efficiency losses or resolution effects are considered. For a more detailed review of available particle generation approaches for MT in Geant4, we refer the reader to \cite{sarah2023}. \\
    
\noindent \textbf{B2G4 data interpreter:} The B2G4 data interpreter module uses the Python extension of ROOT \cite{brun1997root}, PyROOT, to parse simulation output from Geant4 into a data structure suitable for data analysis or machine learning workflows. The ROOT software provides a tree data structure to yield faster access to massive data than regular files. PyROOT provides automatic dynamic bindings between Python and C++, making it ideal for converting the Geant4 output tree data structure into a suitable format for analysis or use by machine learning algorithms.

\section{Muon scattering tomography with B2G4}
\label{sec:muon-tomo-scattering}

\noindent Muon scattering tomography, or MST, estimates the deflection in the muon trajectories as they traverse the VOI. The magnitude of muon deflection, known as the scattering angle $\theta$, supplies information about the material type and density that the muons traverse: materials with higher densities or atomic numbers yield a larger $\theta$ than those with lower densities. Muon trajectories are inferred using tomographic reconstruction techniques to create a 3D scattering density map of the VOI by estimating deflection angles. Hence, for an estimated muon path, the reconstruction algorithms estimate the set of voxels most probably traversed and allocate a score based on their scattering angle. All muon tracks are processed to create a 3D material density map rendering the internal object structure. B2G4 has a module for muon tomography, assuming MST simulations and data interpretation have been carried out. The \textit{B2G4 muon tomography} module takes processed muon hitmaps as input and produces the 3D reconstructed scene designed in Blender as output. B2G4 implements two standard reconstruction algorithms in MST: Point of Closest Approach (PoCA) and Angular Scattering Reconstruction (ASR).\\

\noindent The PoCA algorithm assumes a singular point of scattering for a muon and calculates its minimal separation between incident ($in$) and emergent trajectories ($out$). PoCA employs the incident measured muon path data in terms of position $\mathbf{p}$ and directions $\mathbf{d}$ as $r_{in} = (\mathbf{p_{in}}, \mathbf{d_{in}})$ and $r_{out} = (\mathbf{p_{out}}, \mathbf{d_{out}})$. Therefore, to determine the exact location of the muon deflection, PoCA solves a system of equations given by the parametric equation of points on straight lines, $P=\mathbf{p}+t\mathbf{d}$, being $t$, the variation parameter. Formally, this system of equations can be written as:
\begin{equation}
\left\{
\begin{aligned}
&\phantom{}P_{in} = \mathbf{p_{in}} + t_{in} \mathbf{d_{in}} \\
&\phantom{}P_{out} = \mathbf{p_{out}} + t_{out} \mathbf{d_{out}} \\
&\overrightarrow{P_{in}P_{out}} \cdot \mathbf{d_{in}} = 0 \\
&\overrightarrow{P_{in}P_{out}} \cdot \mathbf{d_{out}} = 0 
\end{aligned}
\right.
\label{eq:poca}
\end{equation}

\noindent Once the PoCA point from both trajectories has been estimated, the muon path is constructed, and voxels crossed by this path will be considered possible candidates influencing the muon trajectory. Then, an information signal (or score) for the muon path, noted as $s$, proportional to the scattering angle $\theta$, is assigned to the voxel associated with the estimated PoCA point. B2G4 applies noise reduction by rejecting extreme angles, then computes the median for each voxel to generate the 3D scattering density map for all muon tracks. While PoCA is efficient and fast, it oversimplifies the 3D representation by assuming single-point scattering and thus neglects scattering across adjacent voxels. Conversely, ASR assigns values to neighboring voxels by defining a distance $D_{r}$, between the center of a voxel $c$ and the incident/emergent muon trajectories::
\begin{equation}
D_r = \max \left( \min_{t_{in}} \left| \left| P_{in}(t_{in}) - c \right| \right|\,,\, \min_{t_{out}} \left| \left| P_{out}(t_{out}) - c \right| \right| \right)
\end{equation}

\noindent with $P_{in}$ and $P_{out}$ the parametric muon paths from equation~\ref{eq:poca}. The value of $D_r$ is compared with a threshold distance $d$: voxels that fulfill the condition $D_{r}<d$ are activated. The threshold value $d$ ensures that at least one voxel for each incoming and emergent track pair is activated as a potential candidate affecting the muon trajectory. The resulting 3D scattering density maps in ASR incorporate contextual information from neighboring voxels, providing a more refined and robust representation compared to PoCA. \\

\noindent Figure~\ref{fig:figure4} illustrates a simplified design of a steel cargo container ($G4\_STAINLESS-STEEL$) with a leaden monkey head inside ($G4\_Pb$). The container scene comprises 27,504 triangles and is exported in 0.30 seconds. Figure~\ref{fig:figure4}.(a) showcases a texture-rich model in Blender, displaying the leaden monkey head, while Figure~\ref{fig:figure4}.(b) depicts the Geant4 import. In Figure~\ref{fig:figure4}.(c), a side view of the container's tomographic reconstruction in PoCA (up) and ASR (low) is presented. Last, Figure~\ref{fig:figure4}.(d) offers a top-view perspective reconstructed with PoCA (left) and ASR (right). Both methods show detailed facets of the monkey head, including ears, chin, and overall shape. However, while PoCA introduces noise due to its single-point scattering assumption, ASR offers a noise-resistant reconstruction by comprehensively considering all potential voxel interactions in the muon path. This makes ASR a more robust method for detailed imaging, particularly in environments with complex scattering phenomena. 

\section{Conclusions}
\label{S:conclusions}
Scene design in Geant4 with precise geometries poses a significant challenge, particularly given the need for manual coding of high-variance scenes. This research work introduces Blender-to-Geant4 (B2G4), a holistic workflow to integrate the 3D design capabilities of Blender with the simulation finesse of Geant4. The proposed B2G4 framework streamlines the creation of corrected and annotated 3D scenes that sidestep the traditionally time-consuming scene generation process. Although B2G4 paves the way toward synthetic data generation and fast scene design, detector creation in Blender and acceleration techniques for solid importation are left for future work. We demonstrate the capabilities of B2G4 in the context of muon tomography for the imaging of concealed objects in a steel container. A specific module in B2G4 implements the reconstruction algorithms to convert simulated data into meaningful 3D scattering maps suitable for further analysis. The created scenes in B2G4 are adaptable for integration into other frameworks, whether in the context of muon tomography or other scientific disciplines that require Geant4 simulations.  
\section*{Conflict of Interest}
\label{SS:ack}
The authors declare that there are no conflicts of interest regarding the publication of this article.
\bibliographystyle{unsrt}
\bibliography{JAIS_template}
\end{document}